\documentstyle[preprint,aps,eqsecnum]{revtex}
\begin{document}
\preprint{December 1993}
\title
{Bosonization of current-current interactions}

\author{D.V.Khveshchenko}
\address
{Physics Department, Princeton University,\\ Princeton, NJ 08544\\
and\\
Landau Institute for Theoretical Physics,\\
2,st.Kosygina, 117940, Moscow, Russia}
\maketitle

\begin{abstract}
\noindent
We discuss a generalization of the conventional
 bosonization procedure
to the case of current-current interactions which get their natural
representation in terms of current instead of fermion number
density  operators.
A consistent bosonization procedure requires a geometrical
quantization of the hamiltonian action of $W_\infty$
on its coadjoint orbits.
An integrable example of a nontrivial
realization of this symmetry is presented by the Calogero-Sutherland
 model.
For an illustrative nonintegrable example
we consider transverse gauge interactions
and calculate the fermion  Green function.
\end{abstract}
\pagebreak

\section{Generalized  bosonization based on $W_\infty$ }
The method of bosonization \cite{LM} proved to be a
 powerful tool for studing a great
variety of one dimensional systems of interacting fermions.
However the applicability of the conventional bosonization technique
is restricted by such necessary requirements as a linear fermion dispersion
 and a locality of a four-fermion interaction which
can be written solely in terms of
density operators $\rho_{\alpha}(q)=\sum_{\alpha p>0}
\Psi^{\dagger}_{\alpha}(p+q)\Psi_{\alpha}(p)$,where
the subscript $\alpha=(R,L)$ labels chirality.
Due to the commutation relations
\begin{equation}
[\rho_{\alpha}(q),\rho_{\alpha^{\prime}}(q^{\prime})]=
\alpha q\delta_{\alpha \alpha^{\prime}}\delta(q+q^{\prime})
\end{equation}
every Hamiltonian bilinear in $\rho_{\alpha}(q)$ can be represented as a
 quadratic form in terms of  free bosons.
 Then the solution of the model can be easily
achieved by means of the Bogoliubov transformation.
It is supposed to yield an asymptotic longwavelength description
of a wide class of four-fermion interactions including long ranged ones.

In a wider sense a
bosonization procedure can be understood as a mapping
of the fermion Hilbert space
 to a space of variables obeying commutation instead of anticommutation
relations. In view of that one might ask whether there exists a
formulation in terms of
some variables of "bosonic" nature which would remain valid at all scales.

Apparently, to perform a consistent bosonization of more general
Hamiltonians which would be  correct
away from the longwavelenth
scaling limit one has to enlarge the algebra of relevant operators.
It naturally stems from the necessity to take into account higher order
spatial
derivatives in Hamiltonians of a general form
\begin{equation}
H=-{1\over{2}}\int dx \Psi^{\dagger}{\partial_{x}^{2}}\Psi+
{1\over{2}}\int dx_1 dx_2 dx_3 dx_4
\Psi^{\dagger}(x_1)\Psi(x_2)V(x_1, x_2; x_3, x_4)
\Psi(x_3)^{\dagger}\Psi(x_4)
\end{equation}
Since the Hamiltonian
(1.2) conserves the number of particles one can construct
a required set of "bosonic"
operators from various fermion bilinears
$\Psi(x)^{\dagger}\Psi(x^{\prime})$.
According to \cite{DDMW} one can choose the following basis of operators
\begin{equation}
W(x,q)=\int dr e^{iqr} \Psi^{\dagger}(r+{x\over 2})\Psi(r-{x\over 2})
\end{equation}
It can be readily seen that operators (1.3) obey the commutation relations
\begin{equation}
[W(x,q), W(x^{\prime},q^{\prime})]=2i\sin{1\over 2}(x^{\prime}q-xq^{\prime})
W(x+x^{\prime},q+q^{\prime})
\end{equation}
corresponding to the infinite dimensional
algebra $W_\infty$ which is a quantum
deformation of the classical algebra  $w_\infty$ of area preserving
(symplectic) diffeomorphisms of the plane $(x,q)$.

This algebra
 can be also understood as a
particular limit of the trigonometric form of
the $SU(N)$ algebra at $N\rightarrow
\infty$\cite{F}. Notice that
in the case of a finite chain of length $N$ lattice
counterparts of operators (1.3) do form the $SU(N)$ algebra.

This algebraic structure finds numerous applications which include
 quantum mechanics of fermions on the first Landau level \cite{CTZ},
combinatorics of Laughlin wave functions\cite{K},
topologically massive 2+1 dimensional gauge theories \cite{Kog},
two dimensional turbulence of an incompressible fluid \cite{Z},\cite{S}
and two dimensional gravity \cite{KP}.

In the context of one dimensional classical
(quantum) Hamiltonian dynamics
the abovementioned
two dimensional manifold is realized as a phase space
of a single particle.
Accordingly, phase space volume presering diffeomorphisms
correspond to canonical (unitary in the quantum case) transformations.

The physical meaning of $W$-operators
 can be easily clarified by expanding
$W(x,q)$ into a series
$W(x,q)=\sum_{s=1}^{\infty} W^{s}(q){x^{s-1}\over {(s-1)!}}$.
A simple analysis shows that the label $s$ can be naturally identified with
the conformal spin of the field $W^{s}(q)$.

In the case of a finite chain
the mode index $q={2\pi n\over N}$ runs over $N$ distinct values.
In the case of a finite density of particles
one may distinguish between left and
right moving particles.
The two ("left" and "right") algebras are isomorphic
to each other. Then the algebra of, say, "right"
 fields $W^{s}_{R}(q)$
resulting from
(1.4) gets a nontrivial central extenion
\begin{equation}
[W^s (q),W^{s^{\prime}}(q^{\prime})]=((s-1)q^{\prime}-(s^{\prime}-1)q)
W^{s+s^{\prime}-2}
(q+q^{\prime})+...+\delta(q+q^{\prime})\delta_{ss^{\prime}}c(q)
\end{equation}
where dots denote contributions of fields with
spins ranging from $s+s^{\prime}-4$ to zero.
The c-number term providing a central extension
to the algebra (1.5) appears in the same way as a conventional
Schwinger anomaly in the r.h.s. of (1.1).
For a formal derivation of the c-number term in (1.5) one has to
redefine the $W$-generators:
$W^{s}_{R}(q)\rightarrow W^{s}_{R}(q)-\sum_{p>0}(p+{q\over 2})^{s-1}
n(p)$ where $n(p)$ is the Fermi distribution function. A
 total anomaly obtained by a summation over all Fermi points
cancels out.

Note that chiral $W_\infty$ algebras naturally appear in the theory
of edge states on boundaries of QHE droplets, the number of independent
species being equal to the number of
closed boundaries \cite{Stone}.

Now we consider more concretely, the algebra
of "right" currents' and fermion
numbers' densities.
The spin one field $W^1 (q)$ obeys the Abelian Kac-Moody algebra
(1.1) and
can be identified with the fermion density operator $W^1 (q)=\rho_{R}(q)$
while the spin two field $W^2 (q)$ has commutation relations
\begin{equation}
[W^2(q),W^1(q^{\prime})]=-q^{\prime}W^1(q+q^{\prime})
\end{equation}
\begin{equation}
[W^2(q),W^2(q^{\prime})]=(q-q^{\prime})W^2(q+q^{\prime})+{c\over 12}
(k_{F}^2 q-q^3)\delta(q+q^{\prime})
\end{equation}
One can readily recognise
the operators
$W^{2}(q)$  obeying the Virasoro algebra (1.7) as current densities
$\pi_{R}(q)=\sum_{p>0}(p+{q\over 2})\Psi^{\dagger}(p+q)\Psi(p)$.
In a proper normalization the central charge $c$ corresponding to the case of
free fermions is equal to one.

To proceed with a consistent quantization of Hamiltonians expressed
in terms of generators of the infinite algebra (1.5) one can apply
the so-called method of
geometrical quantization \cite{Kir}.
Recently it was shown that this method generalizing a coherent state
representation provides a regular quantization procedure for the case of
affine Lie algebras  \cite{W}.

In the framework of this approach one has to derive a covariant
Lagrangian of the system
in terms of variables taking their values on coadjoint
orbits of the underlying algebra.
An arbitrary element of the orbit $Q$ can be represented by
 the projection operator into some coherent state $|\Psi>$
\begin{eqnarray}
Q=|\Psi><\Psi|=gPg^{\dagger}\nonumber\\
g=\exp(i\int dqdx\phi(x,q)W(x,q))
\end{eqnarray}
where $P=|0><0|$ is a projection operator into a reference state $|0>$
belonging to the orbit.
By construction one always has $Q^2 =Q$ and
the orbit is specified by the condition $trQ=1$ where  the
trace symbol stays for the integral over phase space coordinates:
 $tr=\int dqdx$.
The so-called
symbol of an arbitrary operator ${\hat O}$
 is given by its average over some coherent
state: $<\Psi|{\hat O}|\Psi>=tr(QO)$.

A natural parametrization of the
 coadjoint
orbit of $W_\infty$
can be obtained in terms of the phase space density \cite{DDMW}
\begin{equation}
u(x,p)
=\int drdq e^{irp-ixq}<\Psi|W(r,q)|\Psi>
\end{equation}
In terms of this variable
the Lagrangian acquires the form
\begin{equation}
L=<\Psi|i g^{\dagger}d_{t}g - H|\Psi>=
i\int_{0}^{\infty} d\sigma tr (u\{ {\partial_{\sigma}u},
{\partial_{t} u} \}_{MB}) - tr(HQ)
\end{equation}
where $u(\sigma=0)=u(x,q)$ and $u(\sigma=\infty)=u_0=const$.
Note that the first term in (1.10) depends exclusively on the
value of $u(\sigma;x,q;t)$ on the boundary $\sigma=0$. The use
of the cocycle construction enables one
 to write down the Lagrangian (1.10) in a totally
covariant form.

A definition of the
so-called Moyal bracket ${\{},{\}}_{MB}$ appearing in (1.10)
reflects the complexity of commutation relations (1.4)
\begin{equation}
{\{}A,B{\}}_{MB}=2\sin{1\over 2}(\partial_{x}\partial_{q^{\prime}}-
\partial_{q}\partial_{x^{\prime}})A(x,q)B(x^{\prime},q^{\prime})
|_{x=x^{\prime},q=q^{\prime}}
\end{equation}
Despite of numerous potential applications of the above formalism
a geometry
of $W_\infty$-orbits remains
 quite unknown. Some attempt to get a first insight based
on the finite $N$ orbits of $SU(N)$ was undertaken in \cite{Z}.

However one can easily obtain a local equation of motion for the
phase space density
\begin{equation}
{\partial_t}u+\{ {\delta H\over {\delta u}}, u\}_{MB}=0
\end{equation}
In the case
of free fermions the equilibrium solution of (1.12)
is merely $u_{0}(x,p)=\theta (k_{F}^2
-p^2 )$. To consider small deviations from the equilibrium state
one can choose the parametrization: $u(x,p)=\theta (k_{F}^2(x)
-p^2 )$ which leads to the approximation called
the collective field theory \cite{JS}.
Comparing this parametrization with the most general expansion
\begin{equation}
u(x,p)=\theta (k_{F}^2
-p^2 )+\sum_{\alpha}
\delta(p-\alpha k_{F})\rho_{\alpha}(x)+{\partial_p}\delta(p-
\alpha k_{F})\pi_{\alpha}(x)+...
\end{equation}
we conclude that in the framework of the collective field theory
one can use an approximate identity
\begin{equation}
\pi_{\alpha}(x)\approx(\alpha k_{F}+{1\over {2i}}{\partial_x})\rho_{\alpha}(x)
\end{equation}

\section{Calogero-Sutherland model}
A remarkable example of a nontrivial realization of the $W_\infty$ symmetry
is provided by the famous Calogero-Sutherland
model of one dimensional fermions with long-range
forces
\begin{equation}
H={1\over{2}}\int dx \Psi^{\dagger}(-{\partial_{x}^{2}}+\omega^2 x^2 )\Psi+
{1\over 2}\int dxdx^{\prime}
\Psi^{\dagger}(x)\Psi(x){\lambda(\lambda -1)\over{(x-x^{\prime})^2}}
\Psi^{\dagger}(x^{\prime})\Psi(x^{\prime})
\end{equation}
 (or $V(x)\sim (L\sin{x\over L})^{-2}$ in
the compact case). This model manifests a surprisingly simple solution
which allows to consider the Hamiltonian (2.1) as some kind of a "statistical"
interaction between fermions   (compare
with the dicussion of one dimenional spin models with $1/x^2$-exchange
in \cite{Hald}).

It was first pointed out by Sutherland
\cite{Suth} that every state of the model (2.1) can be represented in the form
of the Jastrow-type wave function
\begin{equation}
\Psi(x_1,...,x_N)=\prod_{i<j}^{N}(x_i-x_j)^{\lambda}P(x_1,...,x_N)
\exp(-{1\over 2}\omega^2 \sum_{i}^{N}x_{i}^{2})
\end{equation}
where $P(x_1,...,x_N)$ is a symmetric polynomial,
the ground state corresponding to \\
$P(\{x_i\})=1$.

It readily follows from the results of the previous studies \cite{Poly},
\cite{Shastry}
 that the Hamiltonian (2.1) can be simply expressd in terms of operators
\begin{equation}
W^{s}_{n}=\int dr  \Psi^{\dagger}(r)(-i{\partial_r}-
(\lambda-1)\int dr^{\prime}
{\Psi^{\dagger}\Psi(r^{\prime})\over {r-r^{\prime}}})^{s-1}r^n \Psi(r)
\end{equation}
where a proper normal ordering is assumed. It turns out that the
 operators (2.3)
provide a nontrivial realization of $W_\infty$.

The necessity to use the "covariant"  derivative
$-i{\partial_r}-
(\lambda-1)\int dr^{\prime}
{\Psi^{\dagger}\Psi(r^{\prime})\over {r-r^{\prime}}}$ instead of a usual one
can be easily
deduced in the first quantized formalism.
In the space of  many-body wavefunctions the covariant
derivative acts as  $\prod_{i<j}^{N}(x_i-x_j)^{\lambda-1}
{\partial_{x_{i}}}\prod_{i<j}^{N}(x_i-x_j)^{1-\lambda}$.
Being applied to any of functions (2.2)
this operator leaves the result in the same set
of functions
while an ordinary derivative has this property for $\lambda=1$ only.

One can also see that all modes
with $n<s-1$  annihilate  the  ground state:
$W^{s}_{n}|0>=0$ and reveal its intrinsic symmetries.
On the other hand, acting by modes $W^{s}_{n}$ with $n>s-1$ on the
ground state which appears to be the
highest weight vector of the $W_\infty$ representation,
one creates all
excited states of the form (2.2): $W^{s}_{n}|0>=|excitation>$.

 Notice that the central charge $c$ appearing in the
commutator of $W^{2}(q)$
is still equal to one and
in the scaling limit the only effect of the $1/x^2$-interaction
is a renormalization of the Fermi velocity $v_{F}\rightarrow v_{F}\lambda$
\cite{Kaw}.

An apparent similarity of the functions (2.2) and the variational
Laughlin
wave functions proposed to describe ground states and low-lying excitations
of the $\nu=\lambda^{-1}$ (for odd integer $\lambda$)
yields a simple explanation of the reported high symmetries of
Laughlin states \cite{K}.

The nonlinear
construction (2.3) visualizes the fact that the model (2.1) describes
 an analogue of a "statistical" interaction and  can be understood
as a one dimensional
 counterpart of the two dimensional anyon model.
In fact, the Hamiltonian (2.1) does result from the anyon Hamiltonian
with a statistical parameter
equal to $\theta=\pi(\lambda-1)$
if all particles are placed on the same line.

It appears to be crucially
important that  like the free fermion case
the Hamiltonian (2.1) can be written as a linear
form in generators (2.3):
\begin{equation}
H=W^{3}_{0}+\omega^2 W^{1}_{0}
\end{equation}
Obviously, the set of mutually
commuting operators $I_{s}=W^{s}_{0}$ all commute with (2.4)
and constitute a complete set of integrals of motion of the Calogero
problem\cite{Poly}.

\section{Current-current interactions in D dimensions}
As an instructive  (nonintegrable)
example of an occurence
of $W$-symmetries
we consider fermion  interactions bilinear in current operators
\begin{equation}
H-\mu N=\sum_{\alpha}(W^{3}_{\alpha}(0)-k_{F}^{2}W^{1}_{\alpha}(0))
+{1\over 2}\sum_{\alpha\beta}
\int dq W^{2}_{\alpha}(q)D_{\alpha\beta}(q)
W^{2}_{\beta}(-q)
\end{equation}
The Hamiltonian (3.1) provides the first conceivable
example demonstrating
a relevance of the subalgebra of $W_\infty$ formed by $W^{1}_{\alpha}(q)$
 and $W^{2}_{\alpha}(q)$.
Indeed the only
bosonic description of (3.1) which remains valid at all scales
can be performed in terms of currents' and not fermion numbers'
densities.
To treat the model (3.1)
one has to carry out the geometrical quantization of the
Virasoro subalgebra of $W_\infty$ similar to that described
in \cite{AS} in the context of the two-dimensional conformal field theory.

 In general, the matrix $D_{\alpha\beta}(q)$ couples
$N>1$ Fermi points all together.
As another generalization allowed by the
Lagrangian formalism one can also consider retarded interactions corresponding
to a frequency dependent vertex  $D_{\alpha\beta}(\omega,q)$.

The equation of motion for the phase space density (1.13) reads as
\begin{eqnarray}
\partial_{t} u_{\alpha}(x,p)=p\partial_{x} u_{\alpha}
(x,p)
+p(k_{F}^{2}{\partial_x}-{\partial^{3}_x})
\int dx^{\prime}D_{\alpha\beta}(x-x^{\prime})
u_{\beta}(x^{\prime},p)+\nonumber\\
+\int dx^{\prime}\int dp^{\prime}pp^{\prime}
D_{\alpha\beta}(x-x^{\prime})u_{\beta}
(x^{\prime},p^{\prime}){\partial_x}u_{\alpha}(x,p)
\end{eqnarray}
This nonlinear equation can be only treated perturbatively.
In  the longwavelenth limit
one can neglect the last term in (3.2) as a higher order gradient correction.
The residual linear equation enables to determine a propagator
of the field $u_{\alpha}(q,p;\omega)=\int dxdt e^{iqx-i\omega t}
 u_{\alpha}(x,p;t)$:
\begin{equation}
<u_{\alpha}(q,p;\omega)u_{\alpha}(-q,p;-\omega)>=q(
\omega-qp
+{p\over 12}(k_{F}^{2}q-q^{3}){\hat D}(\omega,q))^{-1}_{\alpha\alpha}
\end{equation}
In accordance with the  previous dicussion
 one obtains a longwavelength limit of
 the expectation value $<\rho_{\alpha}(\omega,q)\rho_{\alpha}(-\omega,-q)>$
simply putting $p$ equal to $k_{F}$ in (3.3).

The formulae (3.2,3.3)
 can be also used for a perturbative
 computation of expectation values of any functionals
 $F\{u_{\alpha}\}$ representing various correlation functions $<W^n (q,t)...
W^m (q^{\prime},t^{\prime})>$.

Furthermore at some circumstances the bosonization scheme
 can be generalized to the case of higher dimensions.
Such a development of a conventional bosonization
of density-density interactions was first proposed in
\cite{L} and recently developed in \cite{H} and  also
discussed in \cite{HM},\cite{CF}.

The basic assumption put forward in \cite{H} is an existence of the Fermi
surface obeying the Luttinger theorem. It can be considered as an extended
object having infinite number of degrees of freedom corresponding
to the Luttinger volume preserving diffeomorphisms.
Fluctuations of the Fermi surface
are associated with the collective modes of the system
(particle-hole excitations)
which
constitute the entire low energy physics. This conjecture is supposed to be
essentially weaker than the statement about
an applicability of the Landau-Fermi liquid theory. Therefore
it may facilitate  an informative
analysis of nontrivial non-Fermi-liquid states.

The key element of the construction are commutation relations of
the D-dimensional analogues of the $W$-generators
\begin{equation}
W^{s}_{\alpha}({\vec q})=\sum_{{\vec p}\in \Lambda_{\alpha}}
(({\vec p}+{{\vec q}\over 2}){\vec n}_{\alpha})^{s-1}
(\Psi^{\dagger}({\vec p}+{\vec q})\Psi({\vec p})-\delta({\vec q})<n({\vec p})>)
\end{equation}
Here the unit vector ${\vec n}_{\alpha}$ is a normal to the Fermi surface
"patch" $\Lambda_{\alpha}$ of the area $S_{D}\sim\Lambda^{D-1}$ ( $\Lambda
<<k_{F}$) centered at the point ${\vec k}_{\alpha}$.
Formally  one should first  construct a proper bosonization scheme for
the case of $N\sim ({k_{F}\over \Lambda})^{D-1}$ coupled Fermi points
and then tend $N$ to infinity.

It was shown in \cite{HM} that a straightforward generalization of (1.1,
1.6,1.7)
for $\rho_{\alpha}({\vec q})=W^{1}_{\alpha}({\vec q})$
and $\pi_{\alpha}({\vec q})=W^{2}_{\alpha}({\vec q})$
\begin{equation}
[\rho_{\alpha}({\vec q}),\rho_{\beta}({\vec q}^{\prime})]=
S_{D}({\vec n}_{\alpha}{\vec q})
\delta_{\alpha\beta}\delta({\vec q}+{\vec q}^{\prime})
\end{equation}
\begin{equation}
[\pi_{\alpha}({\vec q}),\rho_{\beta}({\vec q}^{\prime})]=
({\vec n}_{\alpha}{\vec q})
\delta_{\alpha\beta}\rho_{\alpha}({\vec q}+{\vec q}^{\prime})
\end{equation}
\begin{equation}
[\pi_{\alpha}({\vec q}),\pi_{\beta}
({\vec q}^{\prime})]=\delta_{\alpha\beta}
(({\vec q}-{\vec q}^{\prime}){\vec n}_{\alpha})
\pi_{\alpha}({\vec q}+{\vec q}^{\prime})
+{1\over 12}S_{D}
\delta_{\alpha\beta}\delta({\vec q}+{\vec q}^{\prime})
(({\vec n}_{\alpha}{\vec q})k_{F}^{2}-({\vec n}_{\alpha}{\vec q})^3),
\end{equation}
can be only derived if all momenta lie inside a squat box with
a size $\Lambda_{\parallel}$ along the normal to the Fermi surface
being  much less than a size $\Lambda_{\perp}$ in the  tangent direction.
Otherwise one can not neglect four-fermion terms in the r.h.s. of
(3.5-3.7) and
the above algebra gets spoiled.

In absence of any by-hand-introduced cutoff one might think
that  this  condition can be fulfilled dynamically
if due to the specific features
of the interaction vertex $D_{\alpha\beta}
(\omega,{\vec q})$ the following relations
among transferred energy and momentum hold
\begin{equation}
\omega\sim q_{\parallel}<<q_{\perp}
\end{equation}
Notice that these relations obviously fail in the case of the RPA-screened
long-range density-density interaction $V(q)\sim1/q^\alpha$ with
$\alpha >0$. The vertex dressed by the RPA bubbles
\begin{equation}
D_{\rho\rho}(\omega,q)=
{V(q)\over {1+\Pi_{\rho\rho}
(\omega,q)V(q)}}
\end{equation}
 where $\Pi_{\rho\rho}(\omega,q)\approx 1-q^2/\omega^2$ (at $q<<\omega$)
 is a scalar
polarization operator, develops a pole
characterised by the dispersion $\omega\sim q^{1-\alpha/2}$
at small $q$. Moreover, in the case of Coulomb interaction $(\alpha=2)$
recently considered in \cite{BW}
the collective mode acquires a finite plasmon gap and then is no longer
relevant in the longwavelength limit. Thus an  applicability of the method
to the case of long-range denity-density
interactions remains questionable.

However the conditions (3.8) certainly hold in the case of the
RPA-summed effective
current-current interaction  governed by the transverse vector polarization
$\Pi_{\pi\pi}(\omega,q)=\chi (\omega,q) q^2 +
i\sigma(\omega,q)\omega$.  In the gapless metallic state
$(\chi (\omega,q)\approx const; \sigma(\omega,q)\sim 1/q)$
the effective vertex
\begin{equation}
D_{\pi\pi}^{\mu\nu}(\omega,q)={g^2(\delta^{\mu\nu}-{q^{\mu}q^{\nu}\over q^2})
\over {\chi q^2 +
i\gamma{\omega\over q}}}
\end{equation}
demonstrates
an overdamped pole at $\omega\sim iq^3$.
It was suspected for a long time that the interaction (3.10)
changes drastically  the behavior of fermions with respect to the
free case in both three  \cite{Reiser} and two dimensions \cite{Lee}.

In the longwavelength approximation corresponding to the equality (1.14))
a fermion operator can be represented solely
in terms of
$W_{\alpha}^{1}({\vec q})$:
\begin{eqnarray}
\Psi({\vec r})\sim \sum_{\alpha=1}^{N}
\exp(i{\vec k}_{F}(\alpha){\vec r}+i
\Phi_{\alpha}(r_{\parallel}))O_{\alpha}({\vec r}_{\perp})\nonumber\\
\Phi_{\alpha}({\vec r})=\int {d{\vec q}\over {(2\pi)^{D}}}
e^{i{\vec k}_{F}(\alpha){\vec r}}
{W^{1}_{\alpha} ({\vec q})\over {{\vec n}_{\alpha}{\vec q}}}
\end{eqnarray}
The explicit form  of the ordering operator $O_{\alpha}({\vec r}_{\perp})$
necessary
to maintain anticommutativity of operators at equal $r_\parallel ={\vec r}
{\vec n}$
is strongly dependent on dimension. In particular, in two dimension
one can use the operator
$O_{\alpha}({\vec r})=\exp(i\int d{\vec r}^{\prime}
arg({\vec r}-{\vec r}^{\prime})W^{1}_{\alpha}({\vec r}^{\prime}))$
which is a counterpart of the one-dimensional Jordan-Wigner factor
$O(r)=\exp(i\pi\int dr^{\prime}\Psi^{\dagger}(r^{\prime})\Psi(r^{\prime}))$.

\section{One-particle Green function}
Proceeding along the lines proposed in \cite{H},\cite{HM} and
assuming for simplicity a spherical shape of the Fermi surface one arrives
at the integral representation of the one-particle Green function
\begin{eqnarray}
G({\vec r},t)=<\Psi({\vec r},t)\Psi^{\dagger}({\vec 0},0)>\sim
\sum_{\alpha=1}^{N}\exp(i{\vec k}_{F}(\alpha)
{\vec r}-\nonumber\\
-N\int_{{q}_{\perp}<\Lambda_{\alpha}}
{d\omega d{\vec q}\over{(2\pi)^{D+1}q_\parallel}}
{(1-e^{i{\vec q}{\vec r}-i\omega t})\over{(\omega-
q_{\parallel}+S_{D}
{1\over 12}(k_{F}^{2}q_{\parallel}-q_{\parallel}^{3})
{\hat D}(\omega,{\vec q})+i\delta))_{\alpha\alpha}}})
\end{eqnarray}
Taking the limit of $N\rightarrow \infty$ and keeping the first nonzero term
in the $1/N$-expansion one  obtains the formula
\begin{equation}
G({\vec r},t)\sim\int {d{\vec n}\over {(2\pi)^{D-1}}}
{e^{ik_{F}{\vec n}{\vec r}}\over
{{\vec n}{\vec r}-t+i\delta}}\exp(-\int
{d\omega d{\vec q}\over {(2\pi)^{D+1}}} {(1-e^{i{\vec q}{\vec r}-i\omega t})
D(\omega,{\vec q})\over{(\omega-
{\vec n}{\vec q}+i\delta)^2}})
\end{equation}
where $D(\omega,{\vec q})$ denotes a diagonal matrix element of the  operator
$D_{\alpha\beta}
(\omega,{\vec q})$.

Notice that the formula (4.2) generalizes
the result obtained in \cite{CCM} for the case of $1<D<2$
by the method of "asymptotic Ward identities" based on the conjecture
of a dominant rule
of forward scattering processes. The consideration in \cite{CCM}
was restricted on the case of local interactions.
On the basis of theabove observation we expect
that their analysis could be extended
on the case of long-range interactions in
$D<2+\lambda$, the upper critical dimension being determined  by the exponent
$\lambda$ which
governs the asymptotics $D(\omega,{\vec q})\sim max(\omega;q)
^{-\lambda}$ at small $\omega,q$.

We shall concentrate on the $D=2$ case first.
Then the integral over the transverse transferred momentum $q_{\perp}$
yields
\begin{equation}
\int d^{D-1}q_{\perp}D(\omega,{\vec q})\sim {g^2 \over \omega^{1/3}}
\end{equation}
Other integrations in the exponent give the factor
$\sim\exp(-g^2 max(|t^2 -r_{\parallel}^2 |^{1/6}, r_{\perp}))$
which shows that
the integrand in the residual integral over the Fermi surface
 is strongly peaked at
${\vec n}$ parallel (or antiparallel) to ${\vec r}$.

Calculating the Fourier transform of (4.2)
at energy and momentum close to the (Luttinger)
Fermi surface $(\epsilon=0, p=k_F)$ one obtains the expression
\begin{equation}
G(\epsilon, {\vec p})\sim \int_{0}^{\infty}rdr
\cos k_{F}r \int_{0}^{\infty}tdt
{J_{0}(pr)\over{t^2-r^2+i\delta}}e^{i\epsilon t-g^{2}|t^2-r^2|^{1/6}}
\end{equation}
In particular, at $p=k_{F}$ and $\epsilon
\rightarrow 0$ we recover the asymptotics
\begin{equation}
G(\epsilon)\sim {g^{3/2}\over {\epsilon^{5/4}}}\exp(-
{g^{3}\over {\epsilon^{1/2}}})
\end{equation}
which coincides with the result  of
the eikonal approximation \cite{KS}
and  exhibits a behavior drastically different from the Fermi liquid
one.

In the three dimenional case the integral (4.3) behaves as $\log\omega$
which means that $D=3$ is a critical dimension for the interaction (3.10).
Our simplified consideration leads to the conclusion that the one particle
Green function has Luttinger-type features:
$G(\epsilon)\sim \epsilon^{-1+\eta}$ where $\eta\sim g^2$.
 However we stress that
in contrast to the
$D=2$ result (4.5) the latter estimate can be strongly affected by wasted
terms. A more precise analysis is needed to establish the $D=3$ behavior
reliably.

 One can also obtain correlation functions of more
than two fermion operators \cite{KS} using the same technique.

\section{Conclusions}
We observe that a heuristic attempt to accomplish
a consistent bosonization
of one dimensional fermions with nonlinear dispersion and nonlocal interactions
encounters  such an algebraic structure as a central extension of
$W_\infty$ realized in terms of fermion bilinears.
To proceed with a Lagrangian description one has to find a proper
parametrization of coadjoint orbits of $W_\infty$. The orbit
parametrization
in terms of the phase space density leads to
the generalization of the collective
field theory \cite{JS}. It also provides a regular way to derive
 corrections
to results of the conventional bosonization
due to higher gradient terms which become important away
from the scaling limit.

The  Calogero-Sutherland model
presenting an example of a nonlinear realization of the $W_\infty$
allows a simple construction of integrals of motion in terms of generators
of this algebra.

The bosonization procedure based on $W_\infty$ can be also extended
on  higher dimensions by generalizing the approach of the Ref.\cite{H}.
We discuss  current-current interactions
mediated by a transverse gauge field as a  physically relevant example
where the Virasoro-type subalgebra of $W_\infty$ occurs.
On the basis of this consideration one can
establish the status of the eikonal approximation
earlier applied to this problem \cite{KS}.
It corresponds to the neglect of  higher order gradient terms and
gives essentially the same results as a conventional bosonization
as well as the method of "asymptotic Ward identities" \cite{CCM}.

\section{Acknowledgements}
The author is indebted to F.D.M.Haldane, I.I.Kogan and W.Metzner
for valuable discussions.
This  work was  supported by the NSF Grant.

\end{document}